# Optimizing The Selection of Strangers To Answer Questions in Social Media[1]

JALAL MAHMUD, IBM Research Almaden
MICHELLE ZHOU, IBM Research Almaden
NIMROD MEGIDDO, IBM Research Almaden
JEFFREY NICHOLS, IBM Research Almaden
CLEMENS DREWS, IBM Research Almaden

Millions of people express themselves on public social media, such as Twitter. Through their posts, these people may reveal themselves as potentially valuable sources of information. For example, real-time information about an event might be collected through asking questions of people who tweet about being at the event location. In this paper, we explore how to model and select users to target with questions so as to improve answering performance while managing the load on people who must be asked. We first present a feature-based model that leverages users' exhibited social behavior, including the content of their tweets and social interactions, to characterize their willingness and readiness to respond to questions on Twitter. We then use the model to predict the likelihood for people to answer questions. To support real-world information collection applications, we present an optimization-based approach that selects a proper set of strangers to answer questions while achieving a set of application-dependent objectives, such as achieving a desired number of answers and minimizing the number of questions to be sent. Our cross-validation experiments using multiple real-world data sets demonstrate the effectiveness of our work.

## 1. INTRODUCTION

Hundreds of millions of messages are posted on social media daily, including on networks such as Twitter. Through these messages, users share personal status updates such as "just got married," publicize location-based information such as where they were, where they are presently, or where they plan to go, and voice their sentiment about particular products or services. While the buzz of the crowd on social media provides a great deal of information [Chakrabarti et al. 2011, Sakaki et al. 2010] it is still challenging to obtain certain desired information when needed for two main reasons.

First, the desired information may be in people's head but not yet revealed on social media. Assume that one wants to know the current wait time at a local popular restaurant. Although people who just went to the restaurant have the information, they might not reveal their wait time as part of their status update on social media. One way to address this challenge is request information from social media, but this introduces the second challenge: Even if an information seeker broadcasts their request on social media, the request may not reach a person who has the desired information. This is because broadcasts on social media are typically limited to social networks (e.g., a user's followers on Twitter) and there is no guarantee that any of the people in a seeker's network can satisfy their request. For example, it may be difficult to find a friend on social media who has experience with a particular feature of a digital camera, or who happens to be at a particular restaurant and know its current wait time.

To address the two challenges mentioned above, there are a number of efforts on engaging strangers on social media to obtain information (e.g., Quora[2], Yahoo Answers[3], Facebook

---

[1] This paper is an extended version of [Mahmud et al. 2013]
[2] http://www.quora.com/

Questions[4], Nichols et al. 2012). Such efforts roughly fall into two categories: (1) strangers voluntarily offer desired information, such as providing answers to posted questions (e.g., Quora, Yahoo Answers, Facebook Questions); and (2) questions are directly sent to targeted strangers on social media for answers (Nichols et al. [2012]). As shown in Nichols et al. [2012], this latter effort has a distinct advantage over the first one—the ability to obtain the desired information from the right people at the right time. Unlike the first effort, which passively waits for someone to opt in, the second effort *actively* seeks suitable people who are most likely to have the desired information. For example, people who tweeted about a problem in a software product might be asked to provide information about their experience, the results of which might be used to inform improvements to the product. Not only may it be more likely that people will respond quickly to such a request soon after their experience, but their provided information may be more accurate because the experience is still fresh in their mind. A recent study demonstrates the feasibility of this approach in two domains Nichols et al. [2012].

Although there seems to be value in the second type of the systems, so far there is little research in the space. To capitalize on the potential value of this approach, we are building a crowd-powered, intelligent information collection system, which asks the right strangers at the right time on social media for desired information. Given a user's information request, our system aids the user in collecting the desired information in four steps. First, the system monitors a social media stream (e.g., twitter stream) to identify relevant posts (e.g., tweets generated from foursquare.com indicating people at specific locations). Second, it evaluates the authors of identified posts and recommends a sub-set of people to engage (Figure 1a). Third, it generates questions and then sends them to each selected person (Figure 1b). Fourth, it analyzes received responses and synthesizes the answers together.

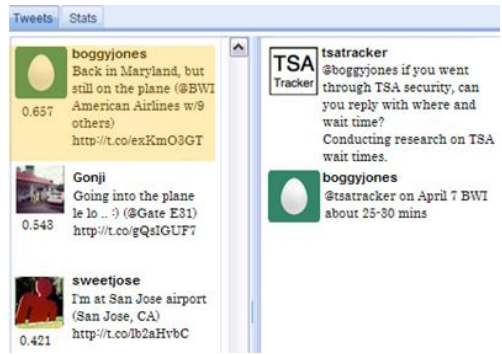

**Figure 1. An example. (a) left panel: three people recommended by our system, and (b) right panel: the question sent and response received.**

In this paper, our focus is on the second step, the process of *automatically* determining the targeted strangers to engage on social media for an information collection task. This process is non-trivial for several reasons. First, the system must determine a person's qualification for the task—whether a person has the *ability* to provide the requested information, such as being at a specific location or having knowledge of a particular product. This requires analyzing the content of one's social media posts, which is often a difficult natural language processing task. Second, the system must determine how *willing* a qualified person is to provide the requested information. We hypothesize that one's willingness to respond to strangers may be related to one's personality traits [Costa et al. 1992], such as openness. However, it is unknown what and how exactly various personal traits influence one's willingness. In addition, accurately predicting one's willingness to respond is critical to the success of our system, since a previous study shows that unwilling people not only would not respond, but are also likely to mark the information request as a spam if they do not understand the request's value, which may result in suspended operations [Nichols et al. 2012]. Third, our system must determine how *ready* a person is to respond at the time of request. As seen later, one's readiness may depend on a number of factors, including the temporal characteristics of one's behavior and their most recent activity on social media. Again, there is little investigation in computationally measuring what and how various factors impact one's readiness to respond.

---

[3] http://www.answers.yahoo.com
[4] http://www.facebook.com

To address the challenges mentioned above, we build computational models to measure a person's *willingness*, *ability*, and *readiness* to respond based on one's behavior on social media. Since measuring one's ability is often domain or context dependent, currently we use a set of domain-specific rules to identify the targeted strangers who might possess the requested information. In the rest of the paper, we focus on the willingness and readiness models, which we believe are mostly domain agnostic. For this reason, throughout this paper we assume that a set of strangers who might know the answer to a question has already been identified, either manually or computationally (e.g., by a set of domain-specific rules). For the sake of concreteness, in this paper we focus on modeling Twitter users, although our work may be adaptable to other social media platforms.

To model one's willingness and readiness to answer questions, we first identify a set of features based on one's social media posts and his/her social interaction behavior. In particular, we hypothesize that one's willingness to answer questions is related to one's personality traits. For example, a friendlier and more extravert person may be more willing to respond to questions posted on social media. Moreover, one's willingness may be reflected by one's past response behavior on social media like Twitter, e.g., the rate of response to others' questions. Likewise, we hypothesize one's readiness to respond to a question is based on a number of features, including the temporal characteristics of one's tweeting behavior and one's most recent tweeting activity. To extract the values of identified features related to willingness and readiness, we analyze the content of one's past tweets and their exhibited tweeting behavior, such as the response time and frequency to others on Twitter. We then train a statistical model to infer the contribution (weight) of each feature to one's willingness and readiness, which are then used to predict one's likelihood to respond to a question.

Even though our statistical model can predict one's likelihood to respond, deciding on whom to send questions is often a nuanced step as different applications or situations may have different goals to achieve. Some examples:

(A) *Obtaining a verifiable answer to a specific question* (e.g., "Can you point to an image on the web showing a black man playing two saxophones in one mouth?"). The goal here is to receive at least one good answer. Once the answer is verified, there is no benefit from additional answers.

(B) *Obtaining multiple opinions* (e.g., quality of a product). The goal of this case is to obtain information from multiple people. However, the benefit of getting additional answers diminishes as the received answers start to converge.

(C) *Obtaining information about a changing situation* (e.g., the wait time at a restaurant). In this case, the goal is to obtain time-sensitive information at defined time intervals (e.g., hourly). Since the number of people who are likely to respond changes over time, there is a need to continuously update the decision on whom to ask. For example, if the system could not find enough people who are highly likely to respond during a particular time interval, it may have to ask people who have a lower likelihood to respond.

As described above, the "cost" of asking questions and the "benefit" from receiving answers depend on the applications. To systematically balance the "cost" and "benefit" in different situations, we develop an optimization algorithm and use it with the statistical model that predicts one's likelihood to respond to select a set of strangers to answer questions. Optimality is measured by a well-defined objective function, for example, maximizing the response rate, or maximizing the expected net benefit from receiving responses.

To validate our approach, we have conducted several cross-validation experiments using real-world question-answer data sets collected from Twitter. The experiments demonstrate the effectiveness of our work in achieving different objective functions. As a result, our work offers two unique contributions:

- A set of features based on one's social media posts and social networking behavior, and a model of a person's willingness and readiness to respond to questions based on those features. The model can be used to predict a user's likelihood to respond to questions on social media.
- A novel approach that combines the above model that predicts a person's likelihood of response with an optimization algorithm to *automatically* select a set of users to whom questions would be sent that optimally achieves application-specific objectives.

The rest of the paper is organized as follows. Section 2 describes previous work related to our research. The subsequent sections describe our experimental system, dataset, baselines, feature extraction, and optimization approach. We describe our experiments in Section 8, offer discussion in Section 9 and conclusions in Section 10.

## 2. RELATED WORK

Our work is most closely related to a recent effort in studying the feasibility of directly asking targeted strangers on social media (Twitter) for information Nichols et al. [2012]. This work studies people's response patterns to questions in two domains and finds that it is feasible to obtain useful information from targeted strangers on social media with an average response rate of 42%. While their findings motivate our work, we advance their manual selection of targeted strangers by developing an *automated* approach that uses a statistical model and an optimization algorithm to select the target strangers on social media while meeting the objectives of a specific application.

Our work is also related to a large number of efforts in online social question-answering (Q&A) services (e.g., WarriorForum[5], Slickdeals[6]). These services rely on people who opt in to provide answers to posted questions. To help get better answers faster, there are a number of efforts on studying the response behavior. For example, Burke et al. study the community responsiveness and the factors that help increase community responsiveness [Burke et al. 2007]. In addition, there is abundant work on identifying expertise [Bouguessa et al. 2008, Pal et al. 2011, Zhang et al. 2007 ]. For example, Pal et al. [2011] uses a machine-learned model to identify experts (or super-users) in the TurboTax Live community. While most of this work models the behavior and traits of potential answerers in terms of their level of expertise, few have modeled an individual's willingness or readiness to answer questions as we do. Moreover, they have not considered how to optimize the selection of potential answerers by balancing a set of cost and benefit factors as in our work.

---

[5] http://www.warriorforum.com/
[6] http://slickdeals.net/forums/forumdisplay.php?f=19

More recently, there is a rich body of work on modeling various aspects of people from social media, including location [Cheng et al. 2010], demographics [Pennacchiotti et al. 2011, Rao et al. 2010], gender [Rao et al. 2010], and political orientation [Pennacchiotti et al. 2011, Rao et al. 2010]. Like this work, we also use one's social media posts and their social networking behavior (e.g., responding to other posts) to model one's traits, including their personality and response behavior. While the existing works focus on modeling the general attributes of people, ours focuses on modeling the attributes that are directly related to one's willingness and readiness to respond to questions on social media.

There is also work on studying and modeling people's responsiveness and availability in various situations (e.g., [Avrahami et al. 2006, Begole et al. 2002, Morris et al. 2010, Paul et al. 2011, Teevan et al. 2011]). Closest to ours is studying response behavior in online social networks, such as Twitter [Morris et al. 2010, Paul et al. 2011, Teevan et al. 2011]. Such effort analyzes question type, response rate, and response time to questions sent *within* one's own social network [Paul et al. 2011, Teevan et al. 2011]. In contrast, our work focuses on identifying and modeling how people respond to *strangers* on social media.

In addition to studying response behavior within one's social network, Avrahami et al. [2006] build statistical models that use a number of features, including "buddy status" and "elapsed time since last message to buddy", to predict one's responsiveness within a certain time interval to incoming instant messages [Avrahami et al. 2006]. Begole et al. [2002] analyze users' desktop activity to find patterns and predict whether a person is available. Like this work, we model similar traits of people (e.g., responsiveness). However, our work makes use of one's social behavior, such as the content of their social media posts and social networking behavior, which poses different technical and analytical challenges compared to the computer usage behavior data used by this previous work.

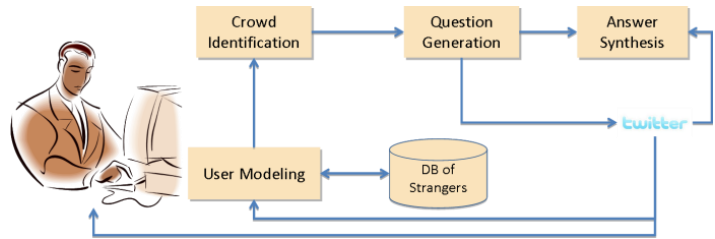

**Figure 2. System Overview**

### 3. OUR SYSTEM
Here we provide an overview of our system and its current applications.

#### 3.1 System Overview
As shown in Figure 2, our system has four key components. The *user modeling* component creates a feature-based profile for each potential target on Twitter based on his/her tweeting activities, including the tweets posted. Each profile contains a set of computed features for modeling one's willingness, ability, and readiness to respond to a request. Such computed profiles are stored in a database. This component also updates the profile of a user (e.g., most recent tweeting activities) who is already in the database. Based on a user's profile, the *crowd identification* component dynamically selects the suitable people to engage with. While a targeted person is identified, the *question generation* component generates a question suitable to the context (e.g., with a proper greeting) and then sends the question to the targeted person via Twitter. The *answer synthesis* component parses and collates the received responses.

Our system can run in one of two modes: (1) manual and (2) auto. In the manual mode, a human operator (e.g., a marketer who wants to engage customers of a brand) watches a Twitter stream and manually selects individuals to send questions, composes the question, and synthesizes the answers received (Figure 3). In the auto mode, the system performs these functions automatically. Our system can also run in a mixed mode, where one component (e.g., crowd selection) runs in the auto mode while another (e.g., question generation) runs manually.

Although our current implementation is based on Twitter, our approach is easily applicable to other social platforms. As long as the system has access to the needed information, such as one's social media posts and social interaction activities, it can then compute needed features and infer one's likelihood to respond to an information request. We already applied our work to our internal enterprise social platform that supports multiple types of social media activities, including facebook-like posts, online communities, and forums, to model and identify potential answerers using their multiple types of social media activities.

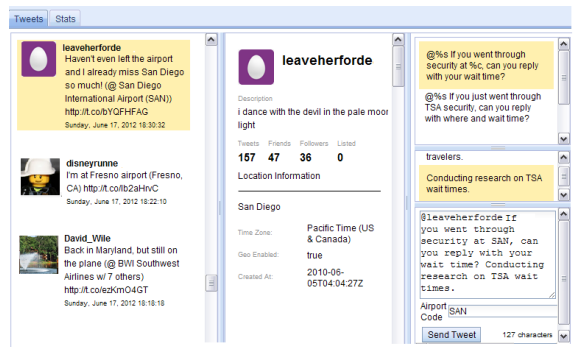

**Figure 3. User interface of our system in the manual mode. The left panel shows a set of filtered tweets (by rules) of those at an airport. The human operator examines the profile of a user and decides whom to engage (middle panel). The right panel allows the operator to compose questions to send.**

### 3.2 Current Applications

Since many people disclose location-based information (e.g., where they are) or their opinions/experience about a product or event on social media, we have built two applications focusing on collecting two types of information on Twitter: location-based information (e.g., airport security check wait time) and people's opinions about a product or service (e.g., cameras, tablets, and food trucks). While each is general enough to represent a class of applications, the two are also distinct enough from each other, which allow us to test the generality of our approach.

Our work should also be easily extended to support other types of information collection tasks, however there may be new challenges that we have not encountered in our current applications. For example, we may need to consider privacy concerns if soliciting potentially sensitive information about people. Moreover, information collection in crisis/emergency management is a key application that we intend to deploy in the near future as the true potential of our work is to collect time-sensitive information from a diverse population.

## 4. DATASET

To build a model for predicting one's likelihood to respond, we ran our system in the manual mode to collect three data sets for both training and testing purposes. Our first two data sets were obtained in the process of collecting location-based information, i.e., airport security check wait time, via Twitter. Specifically, a human operator used our system to manually identify people at an airport in the United States and then ask them for the security check wait time at the airport. Here is an example:

*@bbx If you went through security at JFK, can you reply with your wait time? Info will be used to help other travelers.*

Initially, one human operator sent questions to 589 users and obtained 245 responses (42% response rate), which became our first data set, called *TSA-Tracker*-1. Since we wanted to collect more data in this domain, a different human operator then sent questions to additional 409 users and received 134 responses (33% response rate), which became our second data set, *TSA-Tracker*-2[7]. Our third data set was obtained using the same manual method but by identifying and asking people on Twitter to describe their product/service experience. We asked people's opinions about three types of products: digital cameras, tablet computers, and food trucks. Here are some examples:

*@johnny Trying to learn about tablets...sounds like you have Galaxy Tab 10.1. How fast is it?*

*@bobby Interested in Endless Summer. Sounds like you have eaten there. Is the menu large or small (for a food truck)?*

The questions were sent to 1540 users by one human operator, and 474 responses were received (31% response rate). We refer to this data set as *Product*. In each data set, we also collected the most recent tweets (up to 200 tweets) of each person to whom a question was sent.

## 5. BASELINES

Although the human operators managed to achieve an average response rate of 31-42% through a manual process, the process was found quite difficult as it required the human operator to read the relevant tweets and identify the right people to engage in real time. We thus created two baselines to help assess the general difficulty of the process and evaluate the value of our automated approach. Our first baseline is to engage random people on social media without any filtering. Our second baseline is to let end users (crowd) identify and select targeted strangers to engage on social media.

### 5.1 Asking Random Strangers

To create a true baseline, we sent questions to random people on Twitter without considering their willingness, ability, or readiness to respond to these questions. Specifically, our system *automatically* sent a question to a random person on Twitter at a fixed time interval (e.g., every 5 minutes). We created three different Twitter accounts and experimented with questions on three general topics: weather, public safety, and education. Here is an example question:

*@needy Doing a research about your local public safety. Would you be willing to answer a related question?*

Our aim was to send 250 questions on each topic. However, all three accounts were temporarily suspended by Twitter after sending a certain number of questions. On the weather topic, our system was able to send 187 questions and received only 7 responses. On public safety, 178 questions were sent and only 6 responses were received. On education, only 3 responses were received after sending 101 questions. The response rates on all three topics were very low (well below 5% on each topic). This implies that it is ineffective to ask random strangers on social media without considering their willingness, ability, or readiness to answer questions. Moreover, the account suspension also suggests that many people who received our questions may have flagged our account as a spamming account [Nichols et al. 2012]. Therefore, our work on identifying the right people to engage is critical to the success of our system.

---

[7] We did not combine the two TSA-Tracker data sets since different human operators were involved in the data collection.

**5.2 Crowd as Human Operator**

We determine our second baseline by crowdsourcing the human operator's task and testing a crowd's ability to identify the right targets to engage. We were also curious about the criteria that a crowd would use to identify targeted strangers.

To do so, we conducted two surveys on CrowdFlower, a platform for crowd-sourcing[8]. The first survey asked each participant to predict if a displayed Twitter user would be *willing* to respond to a given question, assuming that the user has the ability to answer. The second survey asked each participant to predict *how soon* the person would respond assuming that s/he is willing to respond. The participant was also required to provide an explanation of their prediction. We deliberately designed two separate surveys to understand a participant's criteria for judging one's willingness and readiness, respectively and to avoid potential interferences.

**5.2.1 Willingness Survey**

The first survey had two settings depending on what information of a Twitter user was displayed to the participant. In the first setting, only the user's tweets were displayed. In the second setting, the user's twitter screen name, twitter profile URL, and tweets were all displayed. For each setting, we randomly picked 200 Twitter users, 100 each from the *TSA-tracker-1* and *Product* dataset. In each setting, we recruited 100 participants on CrowdFlower, each of which was given 2 randomly selected users for prediction. We computed correctness of prediction by comparing participant's predicted response with actual response (responded/not-responded). In the first setting (displaying tweets only), our participants were 29% correct in their prediction, resulting in a 29% response rate if the questions were actually sent. In the second setting, they were 38% correct. Since everything was the same in both settings except the amount of information disclosed about a potential target, the additional information displayed in the second setting helped the prediction task. These results suggest that it is also difficult for the crowd to identify the targeted strangers to engage. We examined the comments by the participants who made correct predictions and found that they relied on several types of information in their prediction.

Most (56.7%) mentioned that they made their predictions by observing users' past responsiveness and interaction behavior: *"Talks to others and responds", and "The user seems extremely social, both asking questions and replying to others"*.

Some participants (10.45%) used one's activeness on Twitter as a predictor: "*This user tweets a lot, seems very chatty"*. Some participants (10.45%) used one's profile information in prediction: *"The user's job is the Social Media representative for H&R block, his job involves answering questions"*; and *"being a social media guy and his tagline saying "we should hang out"*.

Several (6%) also used one's re-tweeting behavior to help them predict: "*No (the person would not respond). Most of the tweets are retweets instead of anything personal"*. It is interesting to note that a small portion were able to infer one's personality (7.4%) and use it for prediction: *"I think

---
[8] http://crowdflower.com/

*he won't respond. Doesn't seem to be very friendly".* However, it was unclear from the responses how the personality was inferred.

We also examined how often the above types of information were used by participants whose predictions were *incorrect*. We obtained similar results: users' past responsiveness (identified by 41% of such participants), one's activeness (7%), profile information (6%), re-tweeting behavior (4%), and personality (3%), respectively. Since the participants considered a similar set of factors regardless their prediction results, this suggests that none of the factors alone is perfect for identifying a user as a responder.

### 5.1.1 Readiness Survey

Our second survey asked each participant to judge how *soon* a person would respond to an information request, assuming that the person would respond. We used a multiple choice question with varied time windows as potential answers (e.g., within an hour and within a day). In this survey, we randomly chose 100 users (responders) from our collected data sets (50 from *TSA-Tracker-1* and 50 from *Product*). We recruited 50 participants on CrowdFlower, each of which was given two randomly chosen responders and their twitter handlers. As a result, 45% (45% of people or users) of the responses indicated that the responders would respond soon (i.e., within an hour), while 55% thought otherwise. We computed correctness of response by comparing participants predicted response within a time window (e.g., within an hour) with actual response within that time window (responded/not-responded within an hour). Thus, if a participant predicted that user "B" will respond within an hour, however, s/he actually did not respond within that time, then the prediction is incorrect. Our participants were 58% correct, which we found by comparing their predicted response times with the actual response times of those users from our data sets. Such a comparison may not be completely fair, since the responses were received at a different time than when the participants examined the accounts. Through their comments, the participants identified several factors that helped them in their tasks. 25% thought that one's activeness (frequent users) and steadiness (consistent usage patterns) on Twitter were good indicators. 30% of them also considered how *promptly* a person responds to others on Twitter a good predictor.

Overall, the crowd-based experiments were very valuable to us. We have learned that it is difficult for an end user to identify the right targets to engage, which justifies our effort to automate the process by recommending targeted people. In addition, our participants used several criteria, including one's personality and responsiveness, to identify the right targets. These criteria thus provided us with an empirical base (feature categories) to build our computational model.

### 6. FEATURE EXTRACTION

To model one's willingness and readiness to respond to information solicitations on Twitter, we have identified five categories of features: *Responsiveness, Profile, Personality, Activity*, and *Readiness*. These features are mainly derived from the content of one's tweets, twitter profile, and social activities on Twitter (e.g., responsive and retweeting behavior). The first four categories, modeling a user's willingness to respond, were mainly identified by our first survey participants when they were predicting whether a twitter user would respond to questions. The last category, modeling one's readiness to respond, was based on our second survey results but augmented to include additional features.

| Big5 Traits | Lower Level Facets |
|---|---|
| Neuroticism | Anxiety, Anger, Depression, Self-consciousness, Immoderation, vulnerability |
| Extraversion | Friendliness, Gregariousness, Assertiveness, Activity level, Excitement-seeking, Cheerfulness |
| Openness | Imagination, Artistic interests, Emotionality, Adventurousness, Intellect, Liberalism |
| Agreeableness | Trust, Morality, Altruism, Cooperation, Modesty, Sympathy |
| Conscientiousness | Self-efficacy, Orderliness, Dutifulness, Achievement-striving, Self-discipline, Cautiousness |

**Table 1. The Big5 personality traits and lower level facets**

| Top-20 Correlations of Friendliness with LIWC categories |
|---|
| Friends (0.23), Leisure (0.22), 1st Person Pl. (0.22), Family (0.2), Other Refs. (0.18), Up (0.18), Social Processes (0.17), Positive Emotions (0.17), Sexual (0.16), Space (0.16), Physical States (0.15), Home (0.15), Sports (0.15), Motion (0.14), Music (0.14), Inclusive (0.14), Eating (0.14), Time (0.13), Optimism (0.13), Causation (− 0.13) |

**Table 2. Example: Top-20 Correlations between the Big5 lower-level facet Friendliness and LIWC categories ($p < 0.05$)**

| Responsiveness Features | Computation |
|---|---|
| Mean Response Time | Avg(T), T denote previous response times |
| Median Response Time | Med(T), T denote previous response times |
| Mode Response Time | Mod(T), T denote previous response times |
| Max Response Time | Max(T), T denote previous response times |
| Min Response Time | Min(T), T denote previous response times |
| Response Rate | $N_R/N_D$, $N_R$ is the number of the user's responses and $N_D$ is the number of direct questions the user was asked in Twitter. |
| Proactiveness | $N_R/N_I$, $N_R$ is the number of user's responses and $N_I$ is the number of indirect questions the user was asked in Twitter. |

**Table 3. Responsiveness features and their computation**

### 6.1 Responsiveness

One's past response behavior on social media was identified as the top factor for predicting one's likelihood to respond. We have thus identified two key features to measure one's overall responsiveness from his/her previous response behavior on Twitter. One is *response time*, which measures how *quickly* a person has responded to a post directed to him/her on Twitter. Our hypothesis is that the faster the person responds in the past, the more likely the person would respond to a request. The other feature is *proactiveness*, which measures how active a person has been to respond to any posts on Twitter including the ones not directed to him/her. The more proactive a person is, the more likely the person would respond to a request. Table 3 shows the two features and their variants. To compute the value of these features, our system collects each person's interaction history, including one's responses to other individuals on Twitter.

| Readiness Features | Computation |
|---|---|
| Tweeting Likelihood of the Day | $T_D/N$, where $T_D$ is the number of tweets sent by the user on day D and N is the total number of tweets. |
| Tweeting Likelihood of the Hour | $T_H/N$, where $T_H$ is the number of tweets sent by the user on hour H and N is the total number of tweets. |
| Tweeting Steadiness | $1/\sigma$, where $\sigma$ is the standard deviation of the elapsed time between consecutive tweets of users, computed from users' most recent K tweets (where K is set, for example, to 20). |
| Tweeting Inactivity | $T_Q - T_L$, where $T_Q$ is the time the question was sent and $T_L$ is the time the user last tweeted. |

**Table 4. Readiness features and their computation**

**6.2 Profile**

Another factor identified by our survey participants is to use one's Twitter profile to assess the *socialness* of the person. We thus extract a profile-based feature *CountSocialWords*, which is the frequency of a set of words, such as "*talking*" and "*communication*", adopted from the LIWC social process category [Pennebaker et al. 2001]. We also added a set of words related to modern social activities, such as "*tweeting*" and "social network".

**6.3 Personality**

Some of our survey participants mentioned personality being one of the factors for predicting one's likelihood to respond. We hypothesize that one's personality traits, such as friendliness and extraversion, may be related to one's willingness to respond to strangers. Since it is impractical to ask a stranger on social media to take a personality test, we investigate how to derive one's personality from his/her social media behavior. Several researchers have found that word usage in one's writings, such as blogs and essays, can be used to infer one's personality [Fast et al. 2008, Gill et al. 2009, Golbeck et al. 2011, Hirsh et al. 2009, Mairesse et al. 2006, Tausczik et al. 2010]. Inspired by the existing work, we used the Linguistic Inquiry and Word Count (LIWC) dictionary to compute one's personality features [Pennebaker et al. 2001, Tausczik et al. 2010]. LIWC-2001 defines 68 different categories, each of which contains several dozens to hundreds of words [Pennebaker et al. 2001]. For each person, we computed his/her LIWC-based personality feature in each category as follows.

Let $g$ be a LIWC category, $N_g$ denotes the number of occurrences of words in that category in one's tweets and $N$ denotes the total number of words in his/her tweets. A score for category $g$ is then: $N_g/N$.

We exclude one's re-tweets when computing LIWC-based personality features, since re-tweets are content generated by others. Besides LIWC categories, psychologists have developed several personality models. One of the most well-studied and used is the Big5 personality model [Costa et al. 1992, Norman et al. 1963]. It characterizes a person's traits from five aspects (also known as OCEAN): *openness*, *conscientiousness*, *extraversion*, *agreeableness*, and *neuroticism*.

Previous work [Fast et al. 2008, Gill et al. 2009, Golbeck et al. 2011, Hirsh et al. 2009] has revealed correlations between the Big5 personality traits [Costa et al. 1992] and the LIWC-category-based features extracted from text, such as blogs, tweets and essays. More recently, Yarkoni et al. [2010] shows that correlations also exist between LIWC features and lower-level facets of Big5 [Costa et al. 1992]. Motivated by these findings, similar to computing the LIWC features, we computed one's Big5 traits and their lower-level facets based on the word use in one's tweets and used them as additional personality features.

To derive personality scores for each of the Big5 dimensions and their lower-level facets, we use the coefficients of correlation between them and LIWC categories found by Yarkoni et al. [2010]. Specifically, we use a linear combination of LIWC categories (for which correlation was found statistically significant by [Yarkoni et al. 2010]), where correlation coefficients are used as weights.

We extract a total of 103 personality features: 68 LIWC features, 5 Big5 trait features, and 30 Big5 lower-level facets.

### 6.4 Activity

As indicated by our survey participants, one's demonstrated activity pattern on Twitter could signal one's willingness to respond. Intuitively, the more active a user is, the more likely the person will respond to a request. To capture this intuition, we compute two features: the total number of status messages (*MsgCount*) sent, and the number of status messages sent daily (*DailyMsgCount*). These features were also used in prior works [Castillo et al. 2011, Lee et al. 2011] and help us distinguish "sporadic" vs. "steady" activeness. We hypothesize that more "steady" users are more dependable and are more likely to respond when asked.

Several of our participants also noted re-tweeting as a factor to determine one's likelihood to respond. We thus incorporate one's retweeing behavior by two features: retweeting ratio (*RetweetRatio*), the ratio between the total number of retweets, and the total number of tweets sent, and daily retweeting ratio (*DailyRetweetRatio*), the ratio between the total number of retweets and the total number of tweets sent daily. These features help us distinguish users who create their own content vs. those who mostly retweet others' [Boyd et al. 2010, Castillo et al. 2011]. We hypothesize that users who create their own content more often may be more willing to respond to a question, since responding involves content creation.

### 6.5 Readiness

Even if a person is willing to respond to a request, s/he may not be ready to respond at the time of request due to various reasons. For example, the person is in the middle of something and unable to respond immediately; or the person is experiencing a device malfunction (e.g., a dying battery) and unable to respond. The ability to predict one's readiness becomes more important when a request requires time-sensitive responses, e.g., the response to the request for the current waiting time at a restaurant. However, one's readiness is highly context dependent and often difficult to capture as shown in our second survey.

We thus use several features to approximate one's readiness building on the features identified by our survey participants. Table 4 lists all the readiness features and their computation. Our rationale of choosing this set of features is two-fold. First, these features are good indicators of one's readiness from a particular aspect. For example, the value of *Tweeting Inactivity* implies one's unavailability. A larger value suggests either that the person is busy and hence uninterruptible, or that s/he is out of reach. Second, these features are easy and fast to compute based on one's past tweeting activity instead of one's tweet content.

### 6.6 Feature Analysis

In total, we identify 119 features to model one's willingness and readiness to respond to a request on social media. Since we do not expect all features to contribute to the models equally, we performed a series of Chi-square tests with Bonferroni correction to identify statistically significant features that distinguish people who responded from those who did not. Using our *TSA-Tracker*-1 data set, we found 45 significant features (False Discovery Rate was 2.8%). In *TSA-Tracker*-2, we found 13 significant features (False Discovery Rate was 11.2%). For the *Product* data set, 33 features were identified as significant (False Discovery Rate was 4.2%).

Table 5 shows the top-10 significant features with their *p*-values in each of our data sets. Since the significant features varied with data set, we were interested in finding a set of features across our data sets for which people who responded consistently scored higher/lower on average than those who did not respond. We list such features in Table 6. As shown later, we use the results of the feature analysis (Tables 5-6) to select features and use them to train our statistical model. We have also run extensive experiments with various combinations of features to hope to find a combination that has significantly discriminative power. We discovered a combination of four top features: *communication* (a LIWC feature), *past response rate*, *tweeting inactivity,* and *tweeting likelihood of the day*. As shown later, this combination produces very impressive prediction results.

| Dataset | Top ten statistically significant features and p-values |
|---|---|
| TSA-tracker-1 | Response Rate (3.8E-7), Mode Response Time (1.75E-5), Tweet Inactivity (3.06E-5), Negative Emotions (3.82E-5), Cautiousness (5.26E-5), Depression (1.09E-4), Excitement-Seeking (1.09E-4), Intellect (1.20E-4), Communication (1.81E-4), Immoderation (1.81E-4) |
| TSA-tracker-2 | Prepositions (1.38E-4), Past (2.86E-4), Exclusion (3.86E-4), Sensation (5.29E-4), Response Rate (2.71E-3), Space (3.54E-3), Tweeting Steadiness (6.3E-3), Achievement-striving (6.6E-3), Agreeableness (7.85E-3), Altruism (8.58E-3) |
| Product | Mode Response Time (0), Tweet Inactivity (2.03E-89), Anger (3.08E-22), Activity Level (2.91E-18), Depression (4.81E-17), Present (5.89E-16), Cautiousness (8.88E-14), Positive Emotion (1.14E-13), Excitement-Seeking (7.43E-13), Response Rate (9.1E-13) |

**Table 5. Top-10 Statistically Significant Features**

| | Features |
|---|---|
| **Higher Average Score of Responded** | Positive Emotion, Positive Feelings, Communication, Inclusion, Sensation, Friendliness, Activity Level, Gregariousness, Trust, Morality and Excitement-Seeking, Extraversion, Agreeableness, Response Rate, Tweeting Steadiness, Tweeting Likelihood of the Day |
| **Lower Average Score of Responded** | Self, 1$^{St}$ Person Singular, Negative Emotion, Anxiety, Mean Response Time, Tweeting Inactivity |

**Table 6. Consistent Features across Data Sets**

## 7. OPTIMIZATION-BASED APPROACH

Using the features described above, here we describe how to construct statistical models that predict a person's likelihood to respond. In addition, we present optimization methods that use our statistical models to select strangers while satisfying a set of objectives.

### 7.1 Statistical Models

Each of our data sets is partitioned randomly into *K* parts for *K*-fold cross validation. For each person in the training set, we compute features as described in Section 6. The features and people's response information are used to train a statistical model.

Given a person in the test set and his/her features computed from tweets, response behavior, and tweeting activity, our goal is to use the trained model to output a probabilistic score that estimates the likelihood for the person to respond[9].

Formally, let $x_i$ be the feature vector of the *i*th person in a data set, and $y_i$ be the response label, such that if the person responded, $y_i = 1$; otherwise $y_i = 0$. In a simplified model, we assume that there is a unit benefit *B* of receiving an answer and a unit cost *C* of sending a question. Of course, depending on the application, this assumption may not suffice. For example, the benefit of receiving additional answers to the same question may diminish as the number of answers grows.

To build an accurate prediction model, we would like to minimize prediction errors. Depending on the application, it may be more preferable to minimize one type of error instead of the other. For example, in one of our applications, the cost of sending a question may be much smaller than the benefit of receiving an answer. In this case, the goal is to minimize the false negative (e.g., missed people who would respond) rather than the false positive (e.g., selected people who would not respond).

However, most of the classifiers typically do not differentiate the two kinds of misclassification errors: false positive and false negative. By default, the classifier-building algorithm assumes that the designer wishes to minimize the overall misclassification rate. Similarly, a standard regression model does not distinguish between overestimating and underestimating probabilities, which in turn results in different types of misclassification errors.

Given the above considerations, we handle the difference in cost by properly weighing the examples, if the unit cost and unit benefit assumption is adequate. We weigh our training examples as follows. Positive examples (i.e., people who responded) are weighted by $B - C$, whereas negative ones (i.e., those who did not respond) are weighted by *C*. In other words, a false negative error, which is a misclassification of a positive example, incurs a missed-opportunity cost of $B - C$, where a false positive, which is a misclassification of a negative example, incurs a cost of a question *C*. Roughly speaking, the weights modify the distribution of the examples, so that examples with larger weight are counted more than the ones with smaller weights. In a more complex model, where the assumption of unit benefit and unit cost does not suffice, benefit and cost can be related non-linearly. In such scenarios, a positive training example may be weighted by $B_i - C_i$ and negative examples by $C_i$, where $B_i$ and $C_i$ may be determined using the given cost and benefit function, respectively.

We have used two popular approaches, Support Vector Machines (SVM) and Logistic Regression, to predict the probability/score for a person to respond. The algorithms were implemented in the R system[10].

---

[9] Alternatively, our models can produce probability estimates $p(x)$ for feature vectors $x$, where a training or test set can be sorted by the probabilities. These probabilistic estimates will not be used directly but rather calibrated by observing actual response rates in the training sets so we can estimate response rates in the subsets of the test sets.

[10] http://www.r-project.org/

### 7.2 SVM-Based Model

Because our data sets are not large, we were able to work with nonlinear rather than linear models, using the primal rather than the dual formulation of the SVM optimization problem. Thus, we seek to minimize the function: $\frac{1}{2}\|w\|^2 + c \cdot \Sigma_i p_i \xi_i$ subject to $y_i \cdot (x_i^T w + b) + \xi_i \geq 1$ and $\xi_i \geq 0$ for every $i$. In this optimization problem, the inputs are as follows. $x_i$ is the feature vector of the $i$th example (including products of original features when we use a quadratic model), $y_i$ is the class label of the $i$th example (either 1 or −1), $c$ is a parameter that scales the misclassification penalty, and $p_i$ is the weight of the $i$th example, namely, $p_i = \frac{1}{2}(y_i+1)B - y_i C$. The decision variables are as follows. $w$ is the vector of feature weights, $b$ is an intercept, and $\xi_i$ is a measure of the misclassification error for the $i$th example.

### 7.3 Regression-Based Model

We have also implemented a logistic-regression model, where the probability $p(x)$ that an instance with feature vector $x$ (including products of original features when we use a quadratic model) is a positive case is estimated as $p(x) = 1/[1+\exp(\alpha+\beta^T x)]$. The coefficients $\alpha$ and $\beta$ are the regression unknowns, which are estimated by the maximum-likelihood principle. The likelihood function is defined so as to reflect the different weights of the examples.

### 7.4 Optimization Methods

In our current implementation, for each person in a data set, our trained statistical models yield a score $s(x)$ for the person's feature vector $x$ as a linear function $s(x) = w^T x$ or a quadratic one, $s(x) = x^T Q x + w^T x$, where $w^T$ is the weight vector, $Q$ has a zero diagonal. The score function defines a weak linear order on the set of feature vectors, so a training or test set can be sorted by the scores. We then use such computed scores to achieve each of our optimization objectives.

### 7.5 Maximizing Response Rate.

Our trained statistical model can be used in different ways to recommend a set of people to engage. The simplest approach is to use it as a binary classifier, which simply predicts a person as a *responder* or *non-responder*. The system can then engage those who are classified as responders. Alternatively, the system can rank the people by the probability scores produced by our statistical model and then select the top-$K$ to engage. However, due to the inherent imperfections (prediction errors) in any prediction models, in reality, the predicted top-$K$ people may not necessarily be the best choices in terms of maximizing the overall response rate. To maximize the overall response rate for a given set of candidates, we thus have implemented the following algorithm.

For each person (candidate) $i$ in a data set, our trained statistical models compute a probability $p_i$. A training or test set can be sorted by such probabilities. We first rank people in the training set by the non-decreasing order of the computed probabilities: $\{p_1,\ldots,p_n\}$. We then focus on intervals in this linear order rather than looking at all possible subsets of candidates. The justification for this restriction is that the linear orders generated by the models exhibit a good correlation with response rates. Figure 4 shows that in general the higher rank of an individual is, the higher the average response rate is (response rate achieved by asking all individuals with a higher rank). Here one's rank is determined by his/her likelihood to respond score computed by our statistical model. A person ranked 1000 means his likelihood to respond is greater than or equal to that of other 999 individuals. Figure 5 shows a similar curve when the model generated from a training set is applied to a test set. While both show that the probability to respond helps predict response rates, they also show that the top segment containing the highest ranked candidates may *not* produce the maximal response rate.

Thus, our approach involves selecting an interval [$i, j$] ($1 \leq i < j \leq n$) from the training set, where the corresponding interval subset $\{p_i,\ldots,p_j\}$ produces an overall maximal response rate among all interval subsets. In our selection, we ignore short intervals at the top of the ranking. The rationale is that the variance in the small sets is large. A short interval that appears to be of a high response rate in the training set does not imply that the corresponding interval in the test set also has a high response rate. We also experimented with a restricted choice of intervals; only those that extend to the top, i.e., of the form [$i, n$]. We have observed, however, that at least when we use linear rather than quadratic, restricting to intervals [$i,n$] rather than [$i, j$] produce suboptimal results.

The best subinterval [$i_r, j_r$] in the training defines a corresponding subinterval [$i_s, j_s$] in the test set, based on percentiles. That is, if $m$ is the cardinality of the test set, then $i_s = [(i_r \cdot m)/n]$ and $j_s = [(j_r \cdot m)/n]$. Another possibility is to select from the test set an interval based on the scores, rather than the ranks, of the endpoints of the optimal interval from the training set, namely, $s(x_{ir})$ and $s(x_{jr})$. In our experiments, we have not seen a clear advantage to one over the other.

We can incorporate additional constraints in our optimal interval selection. For example, we can specify the exact size of the interval (as required by our hypothetical survey example), minimum, or maximum size of the interval as constraints. For example, if a minimum size of the interval is specified, our method will ignore intervals that are smaller than the specified minimum.

**7.6 Maximizing Expected Net Benefit.** Another objective, which arises in many practical applications, is to compare the benefit of getting answers with the cost of asking questions. Using a similar method described for maximizing response rate, we can find an optimal interval from the training set that maximizes the expected net benefit.

Let $C(k)$ be the cost of asking $k$ people and $B(l)$ be the benefit of receiving $l$ answers. Thus, the net benefit of receiving $l$ answers by asking $k$ people is $B(l) - C(k)$. Let $P(k, l)$ be the probability of receiving $l$ answers in response to asking $k$ people, whose likelihood to respond is within a certain interval. The expected net benefit is then $\Sigma_l P(k, l) B(l) - C(k)$. The probability $P(k,l)$ can be estimated by a binomial distribution. Suppose that we decide to send questions to $k$ individuals whose likelihood to respond is in a certain interval, where the response rate is estimated to be $p$.

Then, $P(k, l) = \text{Choose}(k, l)\, p^l (1-p)^{k-l}$.
The probabilities $P(k,l)$ can be computed by the recursive formula: $P(k, l) = P(k-1, l-1) \cdot p + P(k, l-1) \cdot (1-p)$.

Finally, given the estimated response rates in various score intervals in the training set, we can pick an interval in the test set, corresponding to the optimal interval that we have found in the training set.

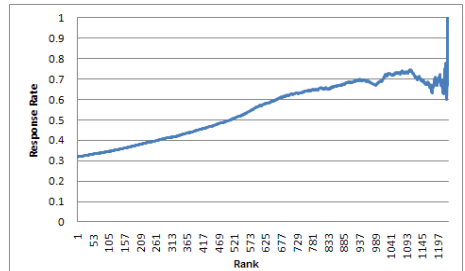

**Figure 4. Response rate in the training set with increasing rank of individuals. A response rate Ri is obtained when (N – i + 1) individuals are asked from the training set in sorted order, where N is the size of the training set.**

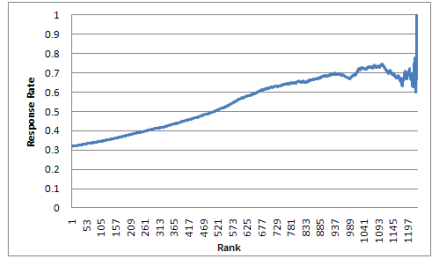

**Figure 5. Response rate in the test set with increasing rank of individuals. A response rate Ri is obtained when (N – i + 1) individuals are asked from the test set in sorted order, where N is the size of the test set.**

### 7.7 Enhancements

We further enhance our method described above to support complex situations in a real-world setting.

**7.7.1 Support of Nonlinear Cost and Benefit Functions**. Our method can support any kind of cost and benefit functions, linear or non-linear. Specifically, we determine a tentative subset of people to ask first by estimating their scores. We then iterate the refinement of the subset-selection optimization. As a result, we calculate, recursively, the expected net benefit for any interval of individuals in the order of their scores to find an optimal interval for any kind of benefit and cost functions.

**7.7.2 Incorporating Additional Constraints**. It is also easy to incorporate additional constraints during the selection process, such as bound by the estimated probability of receiving a certain number of answers. More precisely, when we search for an optimal interval and a number *k*, we ignore those combinations of an interval and number of questions, for which the constraints are not satisfied.

### 8. EXPERIMENTS

To test the performance of our predictive model and recommendation algorithm, we have conducted an extensive set of experiments.

### 8.1 Evaluating Prediction Model

One key piece of our work is to predict the *likelihood* for a stranger on social media to respond to our information requests.

To evaluate the performance of our prediction models, we adopted a set of standard performance metrics including precision, recall, F1, and AUC (Area under ROC curve). For each data set, we performed 5-fold cross validation experiments with uniform weights, where $B – C = C = 1$. Table 7 shows the results for different data sets using SVM and Logistic Regression. Overall, the SVM-based model outperformed logistic regression. The SVM-based model for the *Product* data set also achieved the highest AUC value (0.716). Since the SVM-based model performed better in our experiments, we report various results obtained using this model in the rest of the paper.

|  | TSA-tracker-1 | | TSA-tracker-2 | | Product | |
|---|---|---|---|---|---|---|
|  | SVM | Logistic | SVM | Logistic | SVM | Logistic |
| **Precision** | 0.62 | 0.60 | 0.52 | 0.51 | 0.67 | 0.654 |
| **Recall** | 0.63 | 0.61 | 0.53 | 0.55 | 0.71 | 0.62 |
| **F1** | 0.625 | 0.606 | 0.525 | 0.53 | 0.689 | 0.625 |
| **AUC** | 0.657 | 0.599 | 0.592 | 0.514 | 0.716 | 0.55 |

**Table 7. Performance of prediction models.**

| Minimum Interval Size ($K$) | Optimal Interval | | Training response-rate-optimal | Test response rate-optimal |
|---|---|---|---|---|
|  | Size | [Start, End] | | |
| 0% | 0.5% | [98%, 98.5%] | 1 | 1 |
| 5% | 6% | [91%, 97%] | 0.88 | 0.69 |
| 10% | 11% | [75%, 86%] | 0.82 | 0.67 |
| 20% | 20% | [72%, 92%] | 0.77 | 0.63 |
| 30% | 33% | [61%, 94%] | 0.69 | 0.61 |
| 50% | 51% | [45%, 96%] | 0.60 | 0.55 |
| 70% | 71% | [22%, 93%] | 0.49 | 0.47 |
| 90% | 91% | [9%, 100%] | 0.39 | 0.37 |

**Table 8. Variations of optimal interval, training and test set response-rate with increasing minimum size of the interval (Product data set, SVM model)**

### 8.2 Maximizing Response Rate

In this set of experiments, we demonstrate how our optimization-based approach can be used to maximize the response rate. We used the data sets described in Section 4 for this experiment.

For each data set, we performed 5-fold cross-validation experiments. In our experiment, we have used "asking at least $K\%$ of people" as a constraint to search for the optimal interval that maximizes the response rate. We tested on varied $K$ (e.g., $K$=5%, 10%, etc.) and observed how different optimal sizes were calculated in each case. For all such cases, we assumed uniform weights, where $B - C = C = 1$.

We define the *expected maximum response rate* to be the response rate computed for the optimal interval in the training set. We then computed the response rate on the test set for the same optimal interval and compared it with the expected maximum response rate. Table 8 shows our experimental results for the Product data set using our SVM-based model. As expected, the expected maximum response rate obtained from the training set is higher than the response rate on the test set. Figure 6 shows the variation of the response rate in the test set for our data sets with varied $K$, the minimum number of people (in percentages) to ask.

We see that response rate drops with increasing minimum number of users to ask. Response rates obtained using a SVM-based model were better than the logistic-regression-based model.

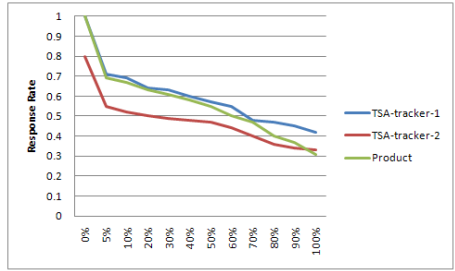

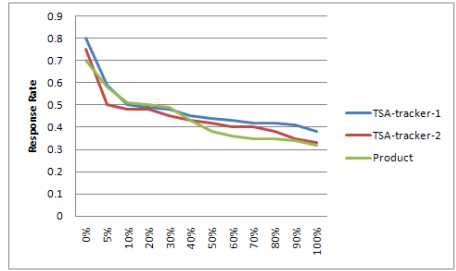

**Figure 6. Response-rate variation on test set with changing minimum percentage to ask – (a) SVM model (b) Regression model**

Regardless which model is used, on average, our approach has improved the response rate across all data sets when compared with baseline (42% for *TSATracker*-1, 33% for *TSATracker*-2 and 31% for Product). On average, when we used an SVM-based model, we got 60% response rates for *TSA-tracker*-1 data set, 47% response rate for *TSA-tracker*-2 data set and 57% response rate for the Product data set. Using logistic-regression-based model, we got 48%, 45% and 44% response rates for these data set.

### 8.2.1 Comparison with different recommendation approaches

For comparison purpose, we also computed the response rates achieved using a simple binary classification (where response rate is the precision of the predictive model) and simply selecting the top-*K* (e.g., *K*=5%, …, 90%. etc.) people by their computed likelihood to respond score. Table 9 shows the results achieved by different approaches. Our algorithm outperformed others in all three data sets.

|  | TSA-tracker-1 | TSA-tracker-2 | Product |
|---|---|---|---|
| Baseline | 42% | 33% | 31% |
| Binary classification | 62% | 52% | 67% |
| Top-K selection | 61% | 54% | 67% |
| Our algorithm | 67% | 56% | 69% |

**Table 9. Comparison of average response rates for different recommendation approaches.**

| Interval Size | Optimal Interval | Response Rate | Recommendation Recall |
|---|---|---|---|
| 25% | [67%, 92%] | 76% | 37% |
| 50% | [46%, 96%] | 68% | 64% |
| 75% | [19%, 94%] | 53% | 82% |
| 100% | [0%, 100%] | 31% | 100% |

**Table 10. Response rate and recall by our recommendation algorithm with fixed size interval (Product Data, all features, SVM Model).**

### 8.2.2 Evaluating Recall of Recommendation

Although our goal is to maximize the overall response rate, we are also interested in finding out what percentage of actual responders that our algorithm identifies. This essentially measures the "recall" of our algorithm. When using a top-$K$ selection, on average, the recall was 57%, 45%, and 63% for the three data sets, respectively. Using our recommendation algorithm, on average over varied $K$ (the minimal percentage of people to ask), the recall was 57%, 46%, and 65%. Since the recall is affected by the number of people are actually engaged, we performed a set of experiments with varied, fixed interval sizes. As shown in Table 10, as the fixed interval size (in percentiles) increases, the response rate decreases and the recall improves. In practice, we may want to find an "optimal" interval size that achieves a balanced high response rate (among the recommended most people would respond) and high recall (most of those who are likely to respond were recommended).

### 8.2.3 Use of Different Feature Sets

We also wanted to see the effect of feature selection on response rates when our optimization method is used. We tried a number of alternatives: personality features alone, responsiveness features alone, readiness features alone, statistically significant features, top-10 statistically significant features (Table 5), top-10 statistical significant features and additional features from products of each pair (10 original and 45 products), consistent features (Table 6) and hand-selected top 4 features (*communication*, *response rate*, *tweeting inactivity* and *tweeting likelihood of the day*) and their products from each pair. Our hand-picked top four features were discovered from an extensive set of experiments. It is interesting to note that these four features are orthogonal to each other: *communication* is a personality feature, *response rate* is a social behavioral feature, and the other two are readiness feature. Table 11 shows response rates for the models obtained using different feature sets when the minimum percentage to ask was set at 5%.

We see a drop of performance when personality, responsiveness, or readiness features were used *alone*. The set of statistically significant features improved the response rates. However, the difference in response rates is quite small when top-ten statistically significant features were used versus all statistically significant features were used. Use of additional features by taking products of each pair of features slightly improved the response rate. The set of consistent features also resulted in small improvements. It is notable that the best response rate for the Product data set was obtained using the top-4 hand-picked features. This feature combination also performed reasonably well for the other data sets.

To understand the overall effect of our approach, Figure 7 shows the comparative performance between our approach and the baseline. Using statistically significant features, on average, our approach produced 67% response rate for *TSA-tracker-1*, 50% response rate for *TSA-tracker-2*, and 62% response rate for Product data set. These results present large improvements over the original response rates at 42%, 33%, and 31%, respectively.

| Feature Set | Response Rate | | |
|---|---|---|---|
| | TSA-tracker-1 | TSA-tracker-2 | Product |
| All | 0.71 | 0.55 | 0.69 |
| Personality | 0.65 | 0.52 | 0.63 |
| Responsiveness | 0.64 | 0.45 | 0.47 |
| Readiness | 0.66 | 0.5 | 0.65 |
| Significant | 0.77 | 0.65 | 0.75 |
| Top-10 Significant | 0.76 | 0.64 | 0.69 |
| Top-10 Significant and their products | 0.77 | 0.65 | 0.74 |
| Consistent | 0.73 | 0.58 | 0.66 |
| Top-4 features and their products | 0.75 | 0.64 | 0.78 |

**Table 11. Response rates using different feature selection, minimum interval was 5%, SVM model.**

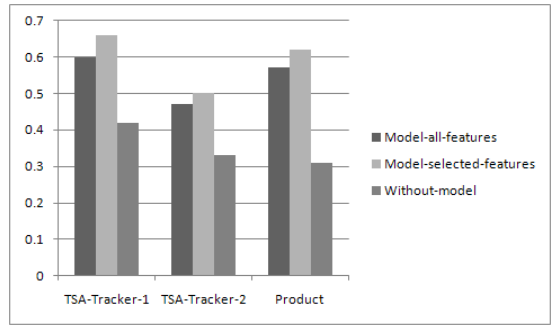

**Figure 7. Improvement of average response rate using response modeling and optimization (SVM model)**

### 8.2.4 Varied Cost and Benefit

So far we have used statistical models that were trained with equally weighed positive (responded) and negative (not-responded) examples, where $B - C = C = 1$, i.e., $B = 2$, $C = 1$. However, depending on the application they may not be uniform. To test how our models perform with varied benefit and cost, we repeated our experiments with different benefit (B) to cost (C) ratio with $C = 1$. We observed varied response rates using our SVM-based model (Table 12). We notice the slight improvement of response rates with the increase of benefit (B) to cost (C) ratio. This result validates our hypothesis that in a system like ours minimizing misclassifications of positive examples (false negative errors) improves prediction accuracy.

| Benefit to Cost Ratio | *TSA-Tracker-1* | *TSA-Tracker-2* | *Product* |
|---|---|---|---|
| 2 | 0.83 | 0.75 | 0.82 |
| 5 | 0.85 | 0.79 | 0.84 |
| 10 | 0.87 | 0.81 | 0.87 |
| 100 | 0.88 | 0.82 | 0.87 |

**Table 12. Response rates using our algorithm for different benefit to cost ratio, minimum interval was 5%, SVM model with significant features.**

### 8.2.5 Domain Sensitivity

To test the domain sensitivity of our model, we built a SVM-based model using statistically significant features from *TSA-Tracker-1* data set. We then applied that to the Product data set and vice versa. The minimum interval size was set at 5%. To make the size of training and test set uniform, we selected a subset of 500 users randomly from each data set. We got 68% response rate for *TSA-Tracker-1* data set and 61% response rate for Product data set. This shows that our models are fairly domain insensitive and can be applied across domains.

### 8.2.6 Live Experiments

In addition to evaluating the effectiveness of our work on previously collected data sets, we also conducted live experiments. Our system was set on the auto mode and dynamically selected strangers on Twitter to send them questions to collect security check wait time at different airports and product experience. In the first case, our system used Twitter's Search API and a set of rules to first identify 500 users who tweeted about being at any US airport. From this set, our system randomly asked 100 users for the security wait time. From the remaining 400 users, our system used the recommendation algorithm to identify 100 users for questioning. It used the SVM-based model with the identified significant features of *TSA-tracker-1*. We waited 48 hours for the responses. 29 of the 100 users from the first set responded (29% response rate) and 66 out of the 100 users from the second set responded (66% response rate).

The same process was repeated for sending product-related questions. The only difference was that the initial set of 500 users was identified using Twitter's Search API and a set of rules to detect users who had tweeted about digital cameras, tablet computers, or food-trucks. Response rates using random question sending was 26% and our algorithm was 60%. Thus, our live experiments also demonstrated the effectiveness of our work in a real world setting.

## 8.3 Maximizing Net Benefit

To test our second objective, we again used our TSA-tracker and Product data sets with 5-fold cross validation. Our optimization algorithm computed an optimal interval from the training set that maximized the net benefit. This optimal interval also defined the corresponding interval in the test set, which selected the subset of people to be asked. We then observed the net benefit that would have been obtained in the test set if that interval were used. We compared this net benefit with the *expected net benefit*—what would have been obtained from the data set if our algorithm was not used. We computed the expected net benefit as follows:

Let $r$ denote the response rate of the data set, and let $M$ denote the total number of people in the test set. If the benefit is $B$ per answer and the cost is $C$ per question (assuming unit benefit and cost), then (without the algorithm) the expected net benefit per question is equal to $rB - C$. If the latter is not positive, then it is best not to send any question. Otherwise, it is best to send questions to everybody in the test set, in which case the expected net benefit is equal to $M$ ($rB -$ $C$). If the algorithm is used, the expected response rate depends on the selected subset of the test set. If the selected subset contains $S$ individuals and its response rate is denoted by $s$, then the expected net benefit from asking only those $S$ individuals is equal to $S(sB - C)$.

Table 13 shows the size of the selected interval, the response rate at the selected interval, the expected net benefit for the interval using our algorithm, and the expected net benefit without using our algorithm for all data sets, when we assumed $B = 10$ and $C = 1$. In each case, an SVM-based model was used. We see that for all of our data sets, the expected net benefit obtained using our algorithm outperformed the expected net benefit without using the algorithm. We also examined how the expected net benefits change when we the ratio of benefit and cost changes.

Specifically, we used different benefit to cost ratio and observed how expected net benefits change. Table 14 shows the result for *TSA-tracker-1* and Product data sets. We observed that, in each of these cases, the expected net benefit obtained using our method outperformed the benefit without using our method. Thus, our method works fairly well to maximize the expected net benefit.

| Data set | Size of optimal interval | Response rate at optimal interval | Expected net benefit (algorithm) | Expected net benefit (without algorithm) |
|---|---|---|---|---|
| TSA-tracker-1 | 66% | 54% | 390 | 377.6 |
| TSA-tracker-2 | 47% | 35% | 236 | 184 |
| Product | 54% | 58% | 724 | 646.8 |

**Table 13. Size of optimal interval, response rate at optimal interval, expected net-benefit with and without algorithm, when, $B = 10$, $C = 1$, and the model is SVM.**

| Benefit-to-Cost ratio | Expected net benefit (TSA-tracker-1) | | Expected net benefit (Product) | |
|---|---|---|---|---|
| | Using our algorithm | Without algorithm | Using our algorithm | Without algorithm |
| 1 | -14.25 | -68.44 | -49 | -212.52 |
| 2 | 19.5 | -18.88 | 32 | -117.04 |
| 5 | 164.8 | 129.8 | 290 | 169.4 |
| 10 | 390 | 377.6 | 724 | 646.8 |
| 100 | 4891 | 4838 | 9291 | 9240 |
| 1000 | 49510 | 49442 | 95734 | 95172 |

**Table 14. Expected net benefits for different ratios of benefit and cost, SVM model was used.**

## 9. DISCUSSIONS
Compared to traditional social Q&A systems, our work offers two unique advantages: *activeness* and *timeliness* in collecting desired information. Our system *actively* seeks out and solicits information from the right people at the *right time* on social media. It is particularly suitable for collecting *accurate* and *up-to-date* information about specific situations (e.g., crowd movement at an airport) from the people who are or just were in the situation. However, the dynamic and complex nature of social media also poses unique challenges for us. Below we discuss several limitations of our current work observed during our investigation and corresponding design implications for building this new class of systems like ours.

### 9.1 Modeling the Fitness of a Stranger to Engage

Currently, we focus on modeling one's likelihood to respond, based on this person's ability, willingness, and readiness to respond to a request on social media. However, our current model does not consider one's traits that may impact the overall *quality* of received responses or cause unexpected viral effects. For example, a more dutiful and trustworthy person may provide a higher quality response, while one's hatefulness might prompt malicious behavior on social media. Moreover, information exchange in our system occurs mainly between the asker (the system) and answerer (a stranger), without any moderation by a larger group. This removes the potential reputation and filtering benefits of typical Social Q&A sites, like Quora, to govern the quality of crowd-sourced information. To ensure the quality of responses and prevent potential mischief, we thus must assess the overall *fitness* of a stranger to engage on social media. Beyond modeling one's likelihood to respond, a fitness model should include people traits that impact various aspects of an information collection process on social media, such as the quality of responses received and potential influence of various response behaviors.

### 9.2 Handling Complex Situations

Our current goal is to recommend the right set of strangers to achieve a particular response rate. While this can handle many real-world applications, it does not cover complex information collection situations. Suppose that our goal is to collect at least *K* data points (e.g., wait times) each hour at a restaurant. In practice, for a given hour, there may be not enough candidates—people who were at the restaurant and tweeted about their location. To handle such situations, we must model and estimate the size of the overall candidate pool in various situations. This may involve estimating the probability of people appearing at the desired location or the probability of publicizing their location on social media. In addition, different applications may have different cost and benefit measures of sending a request and getting a response, respectively. Such benefit or cost may also change as the responses are received. Assume that the goal is to obtain opinions from multiple people. The benefit of getting additional answers diminishes as the received answers start to converge. To systematically handle various constraints, an optimization-based model is then needed to incorporate all the constraints (including benefit and cost) into one objective function, which computes the net benefits. The goal of our recommendation algorithm then becomes to recommend a targeted set of strangers to maximize the net benefits.

### 9.3 Handling Unexpected Answers

Our goal is to ask those people on social media who are most likely to respond to our questions. Due to the public nature of social media, people who are *not* asked can still see our posted questions and may volunteer to provide us with an answer. For example, when collecting our *Product* data set, an owner of a food truck unexpectedly responded to one of our questions. Our current approach has not taken such situations into account. It would certainly be interesting to incorporate such situations into our methodology. This may also become a way to "grow" potential answerers.

### 9.4 Protecting Privacy

Although people volunteer information about themselves on social media, they may not be fully aware how the disclosed information can be used or have complete control over the shared information. We have not conducted in-depth studies on how the people who were selected feel about the process and what their concerns might be. With such information, we can better tune our selection algorithms to exclude people who wish not to be engaged.

## 10. CONCLUSION

As millions of people disclose large amounts of information on social media like Twitter, our current research is on how to best leverage the right people at the right time on social media to crowd-source desired information. In this paper, we have focused on modeling one's willingness and readiness to answer questions based on a set of features extracted from his/her social behavior on Twitter. Using this model, not only can we predict one's probability to respond to a question, but we have also identified a sub-set of features that have significant prediction power. Moreover, we have presented an optimization-based approach that can automatically select a sub-set of candidates to ask based on their likelihood to respond while achieving a set of objectives imposed by specific applications. To validate our work, we conducted a series of cross-validation experiments using multiple real-world data sets. Our experiments demonstrated our approach's effectiveness in satisfying different objectives, including maximizing the response rate and maximize the net benefits.